\documentclass[aps,prc,groupedaddress,showpacs,twocolumn,floatfix]{revtex4}

\usepackage{graphicx}
\usepackage[dvips]{color}
\usepackage{colordvi}

\def\gtorder{\mathrel{\raise.3ex\hbox{$>$}\mkern-14mu
 \lower0.6ex\hbox{$\sim$}}}
\def\ltorder{\mathrel{\raise.3ex\hbox{$<$}\mkern-14mu
 \lower0.6ex\hbox{$\sim$}}}

\newcommand{\rzt}{$\langle r^3 \rangle_{(2)}$}
\newcommand{\rp}{$\langle r^2 \rangle_p^{1/2}$}
\newcommand{\rsq}{$\langle r^2 \rangle_p$}

\newcommand{\be}{\begin{eqnarray}}
\newcommand{\ee}{\end{eqnarray}}
\newcommand{\bes}{\begin{eqnarray*}}
\newcommand{\ees}{\end{eqnarray*}}

\def\bmr{{\bf r}}

\def\bmz{{\bf z}}
\def\minus{\mbox{$-$}}

\begin{document}

\title{Muonic Hydrogen and the Third Zemach Moment}

\author{J. L. Friar}
\affiliation{Theoretical Division, Los Alamos National Laboratory, \\ 
Los Alamos, New Mexico 87545}

\author{Ingo Sick}
\affiliation{Dept. f\"{u}r Physik und Astronomie, Universit\"{a}t Basel, 
Basel, Switzerland}

\date{\today}

\begin{abstract}
We determine the third Zemach moment of hydrogen ($\langle r^3 \rangle_{(2)}$)
using only the world data on elastic electron-proton scattering. This moment
dominates the ${\cal{O}} (Z \alpha)^5$ hadronic correction to the Lamb shift in
muonic atoms. The resulting moment, $\langle r^3 \rangle_{(2)}$ = 2.71(13)
fm$^3$, is somewhat larger than previously inferred values based on models. The
contribution of that moment to the muonic hydrogen 2S level is 
\minus0.0247(12) meV.

\end{abstract}


\maketitle

\section{ Introduction} 
Recent improvements in experimental techniques have led to the
measurement\cite{Niering00} of the 1S-2S interval in hydrogen to an
unprecedented accuracy of 2 parts in $10^{14}$. At the same time the QED
corrections (expressed as powers of $\alpha$, the fine-structure constant) to
this interval have been calculated through ${\cal{O}} (\alpha^7)$, with an
important subset of ${\cal{O}} (\alpha^8)$ terms determined in the past few
years (see Ref.\cite{Eides01} for a comprehensive review of the theory and
Appendix A of \cite{Mohr05} for very recent developments). With these
improvements the difference between experiment and QED theory (including recoil
corrections) is dominated by hadronic size corrections, which affect the 1S-2S
interval in the 10$\underline{th}$ significant figure. Uncalculated QED
corrections probably affect the 12$\underline{th}$ significant
figure\cite{Eides01}.

This situation creates both an opportunity and a problem. The opportunity is to
use the high precision measurements involving S-states to determine the rms
charge radius of the proton at levels of precision of roughly 1\%. Recent values
obtained in this way are: \rp = 0.883(14) fm\cite{Melnikov00}, 0.891(18)
fm\cite{Eides01}, 0.869(12) fm\cite{Pachucki01}, and 0.875(7) fm\cite{Mohr05}.
These consistent results reflect slightly different theoretical and experimental
input.

The murky history of experimental values for the {\em proton} charge radius
obtained from elastic electron-proton scattering data has recently been
clarified by the comprehensive analysis of \cite{Sick03}. That analysis of all
of the world's data separated the charge from the magnetic scattering,
incorporated (significant) Coulomb corrections\cite{Sick96b}, and carefully
treated systematic (as well as random) uncertainties.  The resulting value
of \rp = 0.895(18) fm is significantly higher than most older values, but is
consistent with the atomic determinations. It is unlikely that new and relevant
electron-scattering data will become available in the near future, and
significant improvements over the 2\% uncertainty are therefore unlikely during
this period.

The problem mentioned above is that the extremely precise measurements of
hydrogen spectral lines cannot be used to test QED at anything approaching the
level of accuracy of the QED theory, unless tricks are used to combine
measurements of at least two experiments in such a way that the shortest-range
hadronic and QED processes cancel between the measurements. Although this
complementary approach is a very useful and active field of
research\cite{Karshenboim02}, significant QED information has nevertheless been
removed in the effort to eliminate the leading-order hadronic effects. Any
alternative method that could provide a highly accurate value of the proton
charge radius would resolve much of this problem. A possible method is the
measurement of the 2S-2P Lamb shift in muonic hydrogen, and such an experiment
at PSI \cite{Taqqu99} plans to determine the proton radius.

The Lamb shift in electronic hydrogen is dominated by the (repulsive) radiative
corrections on the electron line, which are much larger than the (attractive)
vacuum polarization corrections on the photon line. The electron spends most of
its time outside the polarization cloud induced in the electron Fermi sea. In
the muonic-atom case the much smaller Bohr radius is within a significant
portion of that cloud and the (electron) vacuum polarization dominates the QED
corrections. The smaller radius also means that the hadronic size corrections
are significantly more important, as well. The goal of the PSI experiment is to
determine the proton charge radius, \rp, to 1 part per thousand. This requires a
theoretical accuracy significantly better than .008 meV, which is the 
uncertainty introduced by a part per thousand error in the proton radius. The
necessary theoretical developments have been recently reviewed in
\cite{Pachucki99}, \cite{Eides01}, and \cite{Borie05}.

Reference \cite{Pachucki99} estimates that uncalculated QED diagrams are likely
to be smaller than 0.002 meV (although Ref.~\cite{Borie05} points out that the
light-by-light-scattering contribution has not yet been calculated). The largest
uncertainties are from hadronic contributions. Most of the latter corrections
fortunately are proportional (or nearly so) to the mean-square charge radius of
the proton. All such terms can be simultaneously fit to the observed muonic 
Lamb shift together with the leading-order size correction in Eqn.~(1) below.
One contribution, however, is significantly different, and estimates of its size
have shown considerable variation. We address this quantity in the next section.

\section{Zemach moments} 

The primary hadronic size corrections are the Coulomb correction of ${\cal{O}}
(Z \alpha)^4$, the one-loop correction of ${\cal{O}} (Z \alpha)^5$, the two-loop
correction of ${\cal{O}} (Z \alpha)^6$, and the hadronic-size modifications of
the radiative and vacuum polarization corrections. In addition there are
hadronic vacuum polarization corrections and polarizability corrections (which
are significant, but which we ignore).

The dominant and long-known\cite{Karplus52} hadronic size correction arises from
modifying the Coulomb potential with the hadronic charge distribution,
$\rho_{\rm ch}$. Higher-order contributions of this mechanism can be obtained as
well, and these were calculated many years ago in the context of muonic
atoms\cite{Friar79,Borisoglebsky79}. The first three orders of corrections for
the $n\underline{\rm th}$ S-state can be written in the general form:
\be \nonumber
\Delta E_n  =  \frac{2 \pi}{3} Z \alpha\, | \phi_n (0)|^2 
\left(  \langle r^2 \rangle -\frac{Z \alpha \mu}{2} \langle r^3 \rangle_{(2)}
\right. \\ 
 +  (Z \alpha)^2 F_{\rm REL} + (Z \alpha \mu)^2 F_{\rm NR} + \cdots \Big)
\, ,
\ee
where $Z$ is the nuclear charge, $\langle r^m \rangle$ is the  $m\underline{\rm
th}$ moment of the nuclear charge distribution (normalized to unit charge),
$\mu$ is the reduced mass of the muon-nucleus system, $\phi_n (0)$ is the muon
wave function at the origin, and the Zemach moment\cite{Zemach56,Friar79}
$\langle r^3 \rangle_{(2)}$ is defined
by
$$
\langle r^m \rangle_{(2)} = \int d^3 r\, r^m\, \rho_{(2)} (r)\, , \eqno (2)
$$
where the convoluted (Zemach) charge density is given by
$$
\rho_{(2)} (r) = \int d^3 z\, \rho_{\rm ch} (|\bmz - \bmr |)\, \rho_{\rm ch} 
(z) \, . \eqno (3)
$$
The nonrelativistic term $F_{\rm NR}$ is part of the Coulomb correction of
relative order (to the leading-order term in Eqn.~(1)) $(Z \alpha )^2 \mu^2
R^2$, where $R$ is a generic proton radius, while the corresponding relativistic
correction is determined by $F_{\rm REL}$, and is of relative order 
$(Z \alpha )^2$.

The finite-size radiative corrections are discussed in detail in
Refs.~\cite{Pachucki99,Eides01,Borie05} and should scale like \rsq. The complete
hadronic correction of ${\cal{O}} (Z \alpha)^6$ has never been worked out but
the Coulomb approximation to it is small (see
Refs.~\cite{Pachucki99,Eides01,Borie05} and above) and this is likely to be
adequate. The remaining term is the one-loop contribution of ${\cal{O}} (Z
\alpha)^5$, discussed in Refs.~\cite{Pachucki96,Pachucki99,Martynenko00,Eides01,
Borie05}, which we discuss next.

The one-loop correction of ${\cal{O}} (Z \alpha)^5$ was considered by
Pachucki\cite{Pachucki96}, and was expressed in terms of the proton's Dirac form
factors, $F_1$ and $F_2$. He also showed in the limit that the proton mass
becomes very large that this contribution approached a relatively simple
expression involving the same limit of the Sachs electric form factor, $G_E
(q^2)$. The expression
$$
\langle r^3 \rangle_{(2)} = \frac{48}{\pi}
\int_0^{\infty} \frac{d q}{q^4} (G_{\rm E}^2 (q^2) - 1 + q^2 \langle r^2 
\rangle_p/3) \, \eqno(4)
$$
can be verified by writing the form factor $G_{\rm E} (q^2)$ as the Fourier
transform of $\rho_{\rm ch} (r)$ and repeatedly integrating by parts in
Eqn.~(4). Substituting Eqn.~(4) for $\langle r^3 \rangle_{(2)}$ in Eqn.~(1)
verifies that Pachucki's simplified expression is just the ${\cal{O}} (Z
\alpha)^5$ term in Eqn.~(1). The terms that vanish in this approximation can be
considered as higher-order recoil corrections, as was noted in
Refs.~\cite{Pachucki99,Eides01}. According to \cite{Eides01} the Coulomb
approximation is good to within 10\% for a simple model of the form factors.

The convoluted density $\rho_{(2)} (r)$ arises because each of the Coulomb
photons is modified at short distances by the nuclear charge distribution. An
alternative description is that the modified Coulomb potential changes the muon
wave function at short distances, and this has an effect on all expectation
values\cite{Zemach56,Friar79,Friar79b}. Thus the $\langle r^3 \rangle_{(2)}$
term above bears the same relationship to the full elastic one-loop contribution
to the Lamb shift as the (traditional Zemach moment) $\langle r \rangle_{(2)}$
term does to the full elastic one-loop contribution to the hyperfine structure.
We also note that analytic results exist for $\langle r^3 \rangle_{(2)}$ for
three simple charge distributions, including the dipole form factor case (viz.,
an exponential charge distribution, for which $\langle r^3 \rangle_{(2)} = 35
\sqrt{3} \langle r^2 \rangle^{3/2}/16$)\cite{Friar79}. The necessity to resort
to models of the proton form factor has lead to significant variations in the
size expected for \rzt. For a treatment using heavy-baryon effective field
theory see Ref.~\cite{Pineda05}.

\section{Calculations}
In Ref.~\cite{Sick03} the {\em world} data on electron-proton scattering for
momentum transfers $q \leq 4$ fm$^{-1}$ have been analyzed (for references to 
the data see \cite{Sick03}).  The electric and magnetic Sachs form factors
$G_E(q^2)$ and $G_M(q^2)$ have been parameterized using a Continued Fraction
(CF) expansion. It has been shown that this CF-expansion is more suitable than
other parameterizations used in the past, and the contribution of the model
dependence due to this choice has been evaluated. The longitudinal/transverse
separation then is done during the global fit of the cross sections, an 
approach that is superior to
the L/T-separations performed when determining $G_E$ and $G_M$  from  
individual data sets.
 
The fit cross sections have been calculated from $G_E$ and $G_M$ including, in
second-order Born approximation \cite{Sick96b}, the Coulomb distortion of the
electron waves; this correction, although neglected in almost all analyses in
the past, is important at low $q$.

The data have been fitted using their random errors, and the error propagation
treated via the error matrix. The systematic uncertainties of the data have been
taken into account by   changing the data sets by the quoted error, refitting
and adding all resulting changes quadratically, hereby obtaining a very
conservative estimate   of the systematic uncertainty of \rzt .

The result: \rzt~ amounts to 2.71(13) fm$^3$, where the error bar includes both
random and systematic errors of the data, the latter dominating by far. For
comparison we quote \rzt~ for the ``standard'' dipole parameterization, which
corresponds to an rms charge radius of 0.811 fm and produces \rzt~ = 2.02
fm$^3$. If the charge radius of the dipole model is scaled to 0.895 fm, the
dipole result becomes 2.72 fm$^3$, which is in very good agreement with the
value we determined directly from the data.

We should perhaps add a comment on the integral in Eqn.~(4), which seems to
indicate that, as a consequence of the $1/q^4$-factor, the \rzt~ depends on data
at extremely small $q$. It must be noted, however, that the $G_E(0)$ term 
cancels the ''1'' at $q\sim 0$, and the $q^2 \langle r^2 \rangle_p/3$ term
cancels the first, $q^2$-dependent term in a power series expansion of $G_E(q)$.
Sensitivity studies have shown that the main contribution to the integral comes
from the region  $q \sim 1.1 \pm 0.5$ fm$^{-1}$ where the data base for
electron-proton scattering is very good.

We have also looked at the effect of two-photon exchange (beyond Coulomb
distortion), which recently has been studied \cite{Blunden03} in connection with
differences between values of $G_E$ extracted from Rosenbluth-separations of
cross sections  and polarization transfer measurements. These corrections turn
out to have a very minor effect; the value of \rzt~ is increased by 0.02 fm$^3$,
{\em i.e.} by only a small fraction of the error bar.

\section{Conclusions} 
We have calculated the third Zemach moment for the charge distribution of
hydrogen from the world's electron-proton scattering data. That moment is \rzt =
2.71(13) fm$^3$, which contributes \minus0.0247(12) meV to the 2S state of
muonic hydrogen. For comparison the result for the ``standard'' dipole model is
an energy shift of \minus0.0185 meV. Our result is model independent, and
removes the concerns of Ref.\cite{Borie05}, who noted that two simple models of
the charge distribution with the same charge radius produced differences of
0.002 meV to the 2S energy shift. Our energy shift is somewhat larger than most
recent values and significantly larger than a few. Our calculation removes most
of the uncertainty from the contribution of the ${\cal{O}} (Z \alpha)^5$
(elastic) finite-size term.

\begin{acknowledgments}

One of us (JLF) would like to thank Edith Borie for
an informative correspondence. The work of JLF was performed under the auspices
of the U. S. Dept. of Energy. The work of IS was supported by the Schweizerische
Nationalfonds.
\end{acknowledgments}

  \end{document}